\documentclass[letterpaper,11pt]{article}
\usepackage{fullpage}
\usepackage{amsmath,amsthm}

\usepackage{times}
\usepackage{soul}

\usepackage[colorlinks=true, citecolor=blue, urlcolor=cyan, linkcolor=purple]{hyperref}
\usepackage{url}
\usepackage[utf8]{inputenc}
\usepackage[small]{caption}
\usepackage{graphicx}

\usepackage{booktabs}
\usepackage[T1]{fontenc}
\usepackage{tgcursor}

\usepackage[maxnames=10,backref=true,style=alphabetic]{biblatex}
\newcommand{\citet}[1]{\citeauthor*{#1}~\cite{#1}}
\newcommand{\citep}[1]{\parencite[#1]}
\usepackage{color}
\usepackage{amsfonts,amssymb,bbm, bm}
\usepackage{empheq}
\usepackage{cleveref}
\usepackage{xcolor}
\usepackage{tcolorbox}
\usepackage{tikz}
\usetikzlibrary{shadows,arrows.meta,positioning,backgrounds,fit,chains,scopes}
\usepackage{caption}
\usepackage{subcaption}
\usepackage{wrapfig}
\usepackage[toc,page,header]{appendix}
\usepackage{minitoc}
\usepackage{multirow}

\usepackage[noend,ruled]{algorithm2e}

\SetCommentSty{mycommfont}
\SetKwInput{KwInput}{Input}
\SetKwInput{KwOutput}{Output}  
\SetAlFnt{\small}
\SetAlCapFnt{\small}
\SetAlCapNameFnt{\small}
\SetAlCapHSkip{0pt}
\IncMargin{-\parindent}
\usepackage{geometry}
\usepackage{amssymb}
\usepackage{array}
\usepackage{graphicx} % For including images
\usepackage{subcaption} % For subfigures
\usepackage{booktabs}
\usepackage[toc, page]{appendix}
\usepackage{cleveref}

\usepackage{enumitem}

\crefname{algocf}{alg.}{algs.}
\Crefname{algocf}{Algorithm}{Algorithms}

\usepackage{cleveref}

\title{SP-Rank: A Dataset for Ranked Preferences with Secondary Information}

\begin{document}

\author{Hadi Hosseini\\  Penn State University, USA\\ \texttt{hadi@psu.edu}
\and Debmalya Mandal\\University of Warwick, UK\\ \texttt{Debmalya.Mandal@warwick.ac.uk}
\and Amrit Puhan\footnote{Authors are ordered alphabetically.}\\ Penn State University, USA\\ \texttt{avp6267@psu.edu}
}

\maketitle
\begin{abstract}
We introduce \textbf{SP-Rank}, the first large-scale, publicly available dataset for benchmarking algorithms that leverage both first-order preferences and second-order predictions in ranking tasks. Each datapoint includes a personal vote (first-order signal) and a meta-prediction of how others will vote (second-order signal), allowing richer modeling than traditional datasets that capture only individual preferences. \texttt{SP-Rank} contains over 12{,}000 human-generated datapoints across three domains—geography, movies, and paintings—and spans nine elicitation formats with varying subset sizes. This structure enables empirical analysis of preference aggregation when expert identities are unknown but presumed to exist, and individual votes represent noisy estimates of a shared ground-truth ranking. We benchmark \texttt{SP-Rank} by comparing traditional aggregation methods that use only first-order votes against SP-Voting, a second-order method that jointly reasons over both signals to infer ground-truth rankings. While \texttt{SP-Rank} also supports models that rely solely on second-order predictions, our benchmarks emphasize the gains from combining both signals. We evaluate performance across three core tasks: (1) full ground-truth rank recovery, (2) subset-level rank recovery, and (3) probabilistic modeling of voter behavior. Results show that incorporating second-order signals substantially improves accuracy over vote-only methods. Beyond social choice, \texttt{SP-Rank} supports downstream applications in learning-to-rank, extracting expert knowledge from noisy crowds, and training reward models in preference-based fine-tuning pipelines. We release the dataset, code, and baseline evaluations (available at \url{https://github.com/amrit19/SP-Rank-Dataset}) to foster research in human preference modeling, aggregation theory, and human-AI alignment.
\end{abstract}

\section{Introduction}

The wisdom of the crowd principle is frequently utilized to recover ground truth rankings for sets of alternatives. Typically, this approach involves aggregating preferences provided by individuals based on their perceptions of the correct answer. However, a fundamental assumption behind this approach, encapsulated by Condorcet's theorem \cite{de2014essai}, is that each individual has a greater than $50\%$ probability of identifying the true ranking. However, when the majority of voters are systematically wrong, such methods fail, as they amplify rather than mitigate collective error.

To address this limitation, \citet{prelec2017solution} introduced the Surprisingly Popular (SP) algorithm, which incorporates not only individual votes (first-order signals) but also meta-predictions—each respondent’s belief about the majority vote (second-order signals). By comparing what individuals think to what they think others believe, the SP algorithm can identify minority expert knowledge and recover the ground truth even when experts are outnumbered. While their original work focused on multiple-choice questions, \citet{hosseini2021surprisingly, hosseinisurprising} extended the framework to more complex ranking settings, including both complete and partial preferences over large sets of alternatives.

Despite these theoretical and algorithmic advances, public datasets that contain second-order predictions remain extremely limited, impeding empirical progress. To fill this gap, we introduce \texttt{SP-Rank}, the first large-scale, publicly available dataset for benchmarking ranking algorithms that exploit both individual preferences and meta-predictions. Each of the \textbf{12,384} human-generated datapoints collected from \textbf{1,152} participants—includes a ranked vote and a meta-prediction of how others would rank the same set of alternatives. The dataset spans three domains (geography, movies, and paintings) and nine elicitation formats with varying subset sizes, enabling controlled study of aggregation across multiple settings.

Using \texttt{SP-Rank}, it is systematically evaluates how incorporating second-order information improves aggregation outcomes over traditional vote-only methods. Our experiments show that SP-Voting consistently outperforms classical aggregation rules—such as Borda, Copeland, and Maximin—not only in recovering full ground-truth rankings but also in recovering local rankings within each subset of alternatives. Furthermore, we demonstrate that SP-Voting remains robust even in elicitation formats with sparse or noisy signals. Finally, we model the structure of the voting population and show that probabilistic models trained on \texttt{SP-Rank} can effectively capture differences between expert and non-expert behavior, revealing both the promise and limits of learning from joint vote-prediction data. Collectively, our findings position \texttt{SP-Rank} as a foundational resource for advancing research in preference aggregation, crowd judgment, and human-AI alignment.

\section{Related Work}

The Surprisingly Popular (SP) framework originated from the Bayesian Truth Serum introduced by \citet{prelec2004bayesian}, which rewards responses that are more common than predicted to encourage truthful reporting. \citet{prelec2017solution} formalized this into the SP algorithm, showing that combining individual answers with meta-predictions enables ground-truth recovery even when the majority is incorrect. For ranking tasks, \citet{hosseini2021surprisingly} showed that even limited second-order information improves rank recovery, while \citet{hosseinisurprising} proposed Aggregated-SP and Partial-SP, scalable variants that reduce elicitation cost and outperform traditional baselines. These contributions were further formalized by \citet{hosseini2025surprisingly}, who analyzed SP under concentric mixtures of Mallows and Plackett–Luce models, providing sample complexity bounds for multi-group populations. The SP framework has since been applied across diverse domains, including incentivizing truthful behavior, eliciting expert knowledge, mitigating biases in peer review, aggregating information, and enhancing predictive accuracy in ensemble methods and social forecasting \cite{prelec2004bayesian, schoenebeck2021wisdom, schoenebeck2023two, kong2018eliciting, lu2024calibrating, chen2023wisdom, rutchick2020does, lee2018testing, luo2023machine, yamashita2025analysis}. Despite this progress, prior work has lacked large-scale datasets with joint vote and prediction data—a gap we fill with \texttt{SP-Rank}.

The challenge of recovering ground truth from noisy individual judgments has been extensively studied in social choice theory \cite{galton1949vox, de2014essai, surowiecki2005wisdom}. A wide range of vote aggregation rules have been proposed to address this, including classical methods such as Borda \cite{borda1781m}, Copeland \cite{copeland1951reasonable}, and Young’s rule \cite{young1977extending}, among others \cite{de2014essai}. These rules aim to aggregate individual preferences into a consensus ranking and can be adapted to handle ranked inputs from voters \cite{boehmer2023rank}. Beyond axiomatic approaches, information aggregation has also been studied from a statistical perspective \cite{de2014essai, conitzer2009preference, xia2010aggregating, conitzer2012common, marden1996analyzing}. However, limited work has addressed aggregation in settings where two layers of information are available—individual votes and meta-predictions of others’ votes. Prior empirical studies of the Surprisingly Popular method have relied on either synthetic simulations or small-scale elicitation, leaving no standardized benchmark for systematic evaluation. To support future research in this direction, our \texttt{SP-Rank} dataset enables direct comparison between traditional vote-based methods and approaches that leverage second-order information, facilitating systematic evaluation across different aggregation strategies.

\section{\texttt{SP-Rank} Dataset}

\begin{figure}[htbp]
    \centering
    \includegraphics[width=0.85\textwidth]{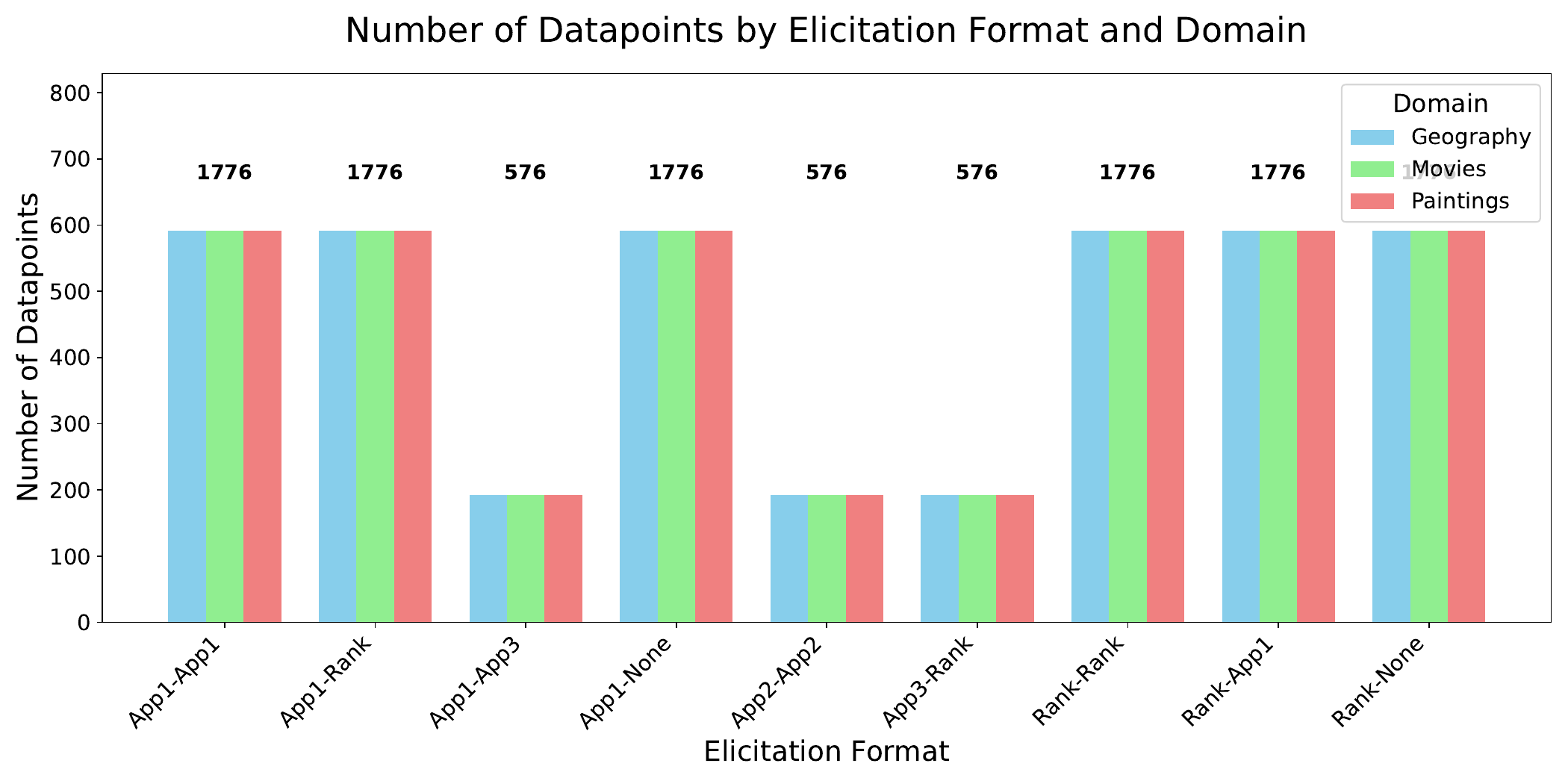}
    \caption{Number of datapoints by domain across each elicitation format. The total number of datapoints is annotated above each format group.}
    \label{fig:datapoints_domain_elicitation}
\end{figure}

The dataset comprises of ranked preference data collected from \textbf{1,152 participants}, each identified by a unique key in the \texttt{workerid} column. For each question, participants provide two types of information: their own \textbf{vote} (personal ranking or selection) and their \textbf{prediction} of how they believe others will vote (meta-prediction). Preferences are collected across three distinct domains: \textbf{Geography}, \textbf{Movies}, and \textbf{Paintings}. Participants are divided into two groups based on the number of alternatives they evaluate: \textbf{720 participants} responded to questions involving \textbf{4 alternatives}, while \textbf{432 participants} responded to questions involving \textbf{5 alternatives}. A sample question that elicits ranked data consists of two parts: Vote - “How do you think the following countries should be ordered from most populated (top) to least populated (bottom)?” and Prediction - “Imagine that other participants will also answer the previous question. In your opinion, what will be the most common ordering of the following countries?” For the 4-alternative condition, a sample set might include: \textit{United States, Russia, Vietnam, United Kingdom}. For the 5-alternative condition, an additional item such as \textit{Kenya} is included: \textit{United States, Russia, Vietnam, United Kingdom, Kenya}.

Each participant interacts with multiple voting problems. Each value under the \texttt{problem} column corresponds to a unique question instance (for example, ranking a set of countries, movies, or paintings) for a particular elicitation format. The \texttt{treatment} column denotes the elicitation format used for that instance, and the \texttt{domain} column specifies the category (Geography, Movies, or Paintings). Thus, each datapoint in the dataset is associated with a unique \texttt{[problem, treatment, domain]} tuple. Each \texttt{workerid} may therefore be associated with multiple datapoints. In total, the dataset comprises \textbf{12,384 datapoints}.

We now describe the elicitation formats, data collection method, and domains in the following subsections in greater detail.

\subsection{Elicitation Formats}

The dataset contains two primary elicitation types:
\begin{enumerate}
    \item \textbf{AppK Elicitation}: Participants select and approve their top \texttt{K} alternatives, without specifying an order.
    \item \textbf{Rank Elicitation}: Participants provide a complete ranking over all given alternatives.
\end{enumerate}

Each datapoint records a \texttt{Vote-Prediction} pair as the elicitation format, consisting of the participant's own \textbf{vote} (their personal ranking or selection) and their \textbf{prediction} of the overall group behavior. A total of \textbf{nine elicitation formats} are implemented, differing in the methods used for voting and prediction. The nine formats are:
\begin{itemize}
    \item \texttt{App1-None}: Participants report their top choice but do not provide any prediction.
    \item \texttt{App1-App1}: Participants report their top choice and predict the population's top choice.
    \item \texttt{App1-App3}: Participants report their top choice and predict the top three choices approved by the population.
    \item \texttt{App1-Rank}: Participants report their top choice and predict a full ranking by the population.
    \item \texttt{App2-App2}: Participants select their top two choices and predict the top two choices approved by the population.
    \item \texttt{App3-Rank}: Participants select their top three choices and predict a full ranking by the population.
    \item \texttt{Rank-None}: Participants provide a full ranking but do not make any prediction.
    \item \texttt{Rank-App1}: Participants provide a full ranking and predict the population's top choice.
    \item \texttt{Rank-Rank}: Participants provide a full ranking and predict the population's full ranking.
\end{itemize}

\Cref{fig:datapoints_domain_elicitation} shows distribution of elicitation format in our dataset. The counts of ApprovalK is lesser than the rest because they only occur where voters have voted over 5 alternatives.

\subsection{Data Collection Method}
We conducted a large-scale empirical study on Amazon Mechanical Turk (MTurk) to gather preferences across different elicitation formats.  Participants provide their vote and prediction information over a set of alternatives. A total of 1,152 participants contributed responses under two experimental conditions distinguished by the subset size of alternatives shown per question—720 participants interacted with subsets of size 4, and 432 participants with subsets of size 5. Despite this difference in subset size, the study design was consistent in its use of three domains as described in \Cref{subsec:domains}. Because no repository of joint vote–prediction data exists, our controlled large-scale collection on MTurk provides a practical and replicable foundation for studying aggregation methods under realistic but tractable conditions.

Participants in both conditions answered questions composed of alternatives sampled from a truncated global ranking: the top 50 items per domain in the 4-alternative condition, and the top 36 in the 5-alternative condition. In each case, alternatives within a subset were selected such that any two adjacent options were separated by six ranks in the underlying ground truth and then randomized before being presented to the voters. This spacing was chosen to balance informativeness with cognitive load, while also ensuring stability across time—particularly in domains like country population—where it is highly unlikely that an item would shift by six or more positions year over year. This design minimizes the chance that minor updates to external data would impact the consistency of ground-truth labels across different iterations of the dataset. The subset size influenced the number of distinct questions we could generate—10 per elicitation format per domain in the 4-alternative case and 12 per elicitation format per domain in the 5-alternative case—while maintaining consistent inter-alternative gaps. To elicit preferences, we used multiple elicitation formats (six in the 4-alternative condition and nine in the 5-alternative condition). Each participant was randomly assigned two formats and answered questions accordingly. In the 4-alternative condition, each participant answered 10 questions (5 per format), while in the 5-alternative condition, each participant answered 12 questions (6 per format). \Cref{fig:datapoints_domain_elicitation} shows distribution of datapoints by elicitation format and domain in our dataset. As shown, the \texttt{App1-None}, \texttt{App1-App1}, \texttt{App1-Rank}, \texttt{Rank-None}, \texttt{Rank-App1}, and \texttt{Rank-Rank} occur in both 4-alternative and 5-alternative conditions but \texttt{App1-App3}, \texttt{App2-App2}, and \texttt{App3-Rank} only occur in the 5-alternative condition.

Additional integrity check included a recall quiz at the end of the survey, where participants were asked to identify their response to a previously answered question. Participant eligibility was restricted in both conditions to MTurk users with at least a 90\% approval rate, over 100 completed tasks, and location restricted to the US East region. Compensation consisted of a \$0.50 base payment for successful quiz completion and a \$0.50 bonus for answering the recall quiz question properly. 

\subsection{Domains}
\label{subsec:domains}
Participants provide report within the following domains - 1) \textbf{Geography} - Participants report countries in decreasing order of their population, \textbf{Movies} - Participants report movies in decreasing order of gross box office lifetime earnings, and, \textbf{Paintings} -Participants report paintings based on decreasing auction prices. Each domain contains multiple questions. Participants are shown a set of alternatives and asked to either provide a \texttt{App1}, \texttt{App2}, \texttt{App3}, or a complete \texttt{Rank} according to the domain-specific instructions. \Cref{fig:datapoints_domain_elicitation} also shows distribution of datapoints across the three domains in our dataset.

\subsection{Voter Accuracy}

\begin{figure}[htbp]
    \centering
    \begin{minipage}[t]{0.49\textwidth}
        \centering
        \includegraphics[width=\textwidth]{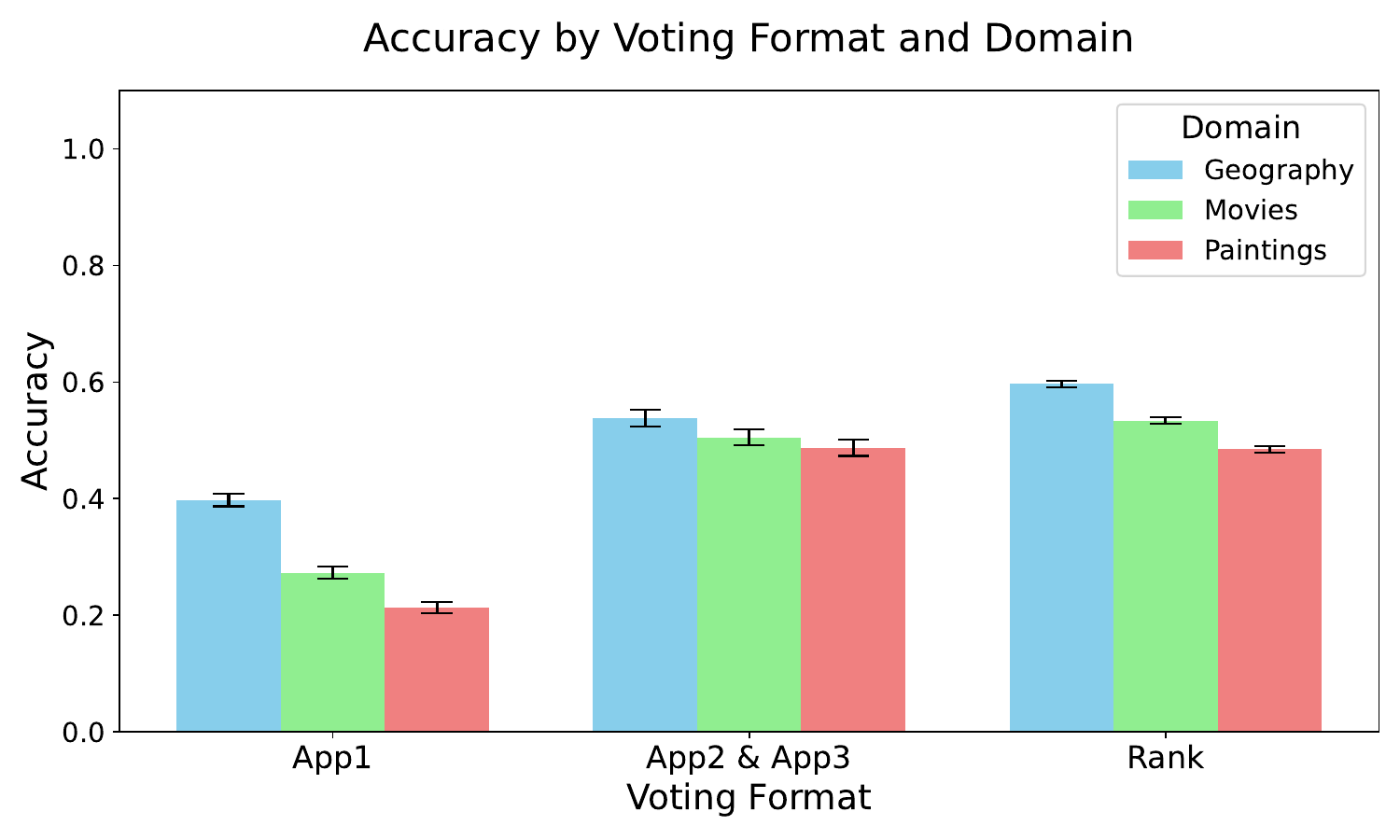}
        \caption{Accuracy by voting format and domain.}
        \label{fig:accuracy_format_domain}
    \end{minipage}
    \hfill
    \begin{minipage}[t]{0.49\textwidth}
        \centering
        \includegraphics[width=\textwidth]{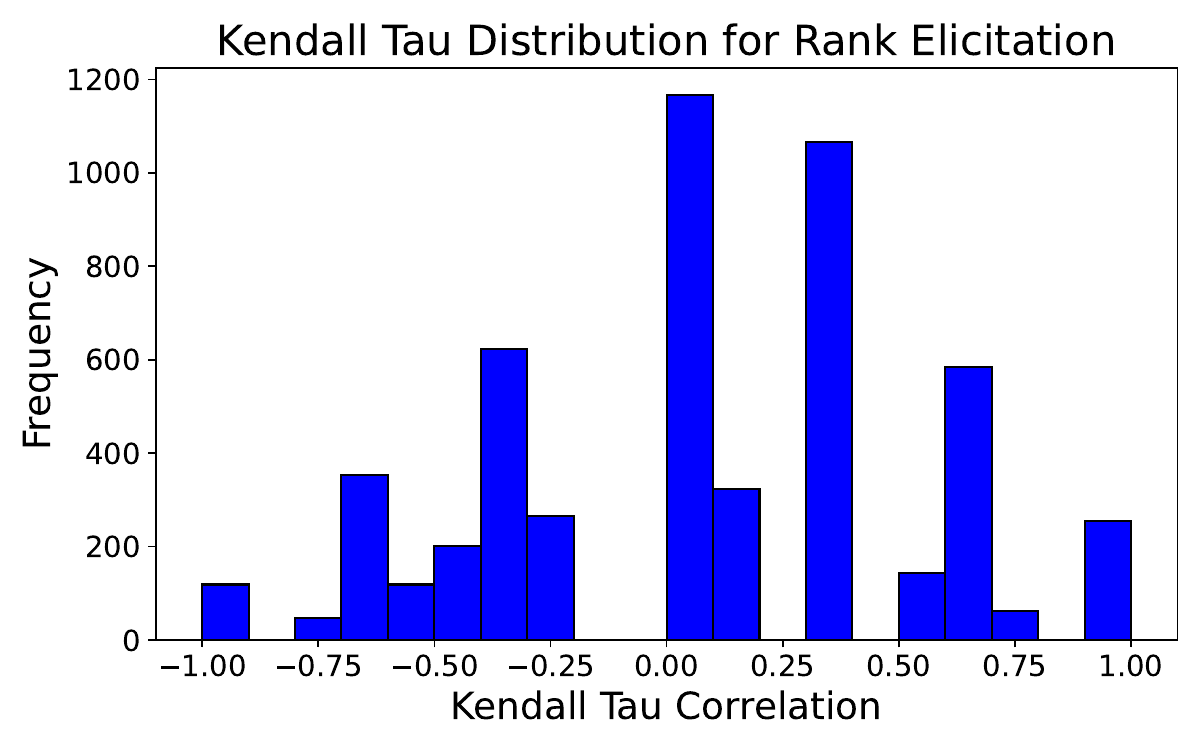}
        \caption{Distribution of Kendall Tau correlations for rank elicitation format.}
        \label{fig:kendall-tau-distribution-rank}
    \end{minipage}
\end{figure}

The Vote part of the Vote-Prediction elicitation format consists of each voter's own opinion. In \texttt{SP-Rank}, the votes were acquired by the following elicitation formats: App1, App2, App3, and Rank. \Cref{fig:accuracy_format_domain} shows the accuracy of the voters in \texttt{SP-Rank} for each domain and elicitation format. The accuracy was computed separately for each elicitation format to assess how closely a participant's vote aligned with the ground-truth.

For the App1 format, where participants selected a single alternative, accuracy was binary:
\[
\text{Accuracy}_{\text{App1}} = 
\begin{cases}
1, & \text{if } a_1 = a_1^* \\
0, & \text{otherwise}
\end{cases}
\]
where \(a_1\) is the selected option and \(a_1^*\) is the top-ranked alternative in the ground-truth ordering.

In the App2 and App3 formats, participants selected two or three alternatives, respectively. Accuracy was computed as the proportion of selected options that appeared in the top-\(k\) ground-truth set:
\[
\text{Accuracy}_{\text{App}k} = \frac{|\{a_i \in A : a_i \in A_k^*\}|}{k}
\]
where \(A\) is the set of selected alternatives and \(A_k^*\) is the set of ground-truth top-\(k\) alternatives.

For the Rank format, participants were asked to rank all presented alternatives. Accuracy was computed using Kendall’s Tau correlation between the participant’s ranking and the ground-truth ranking, i.e., \texttt{kendall\_tau(voter\_rank, ground\_truth)}. \Cref{fig:kendall-tau-distribution-rank} shows the full distribution of Kendall’s Tau correlation values for all the votes. The Kendall’s Tau correlation coefficient is a measure of the ordinal association between two rankings and is given by:
\[ \tau(\sigma', \sigma^*) = \frac{2}{n(n-1)} \sum_{i < j} \mathbf{1}((\sigma'(i) - \sigma'(j))(\sigma^*(i) - \sigma^*(j)) > 0) - 1 \]
where \(n\) is the number of elements in the ranking and $\sigma'$ represents voter\_rank and $\sigma^*$ represents ground\_truth.

Understanding the accuracy distribution across elicitation formats using only individual votes provides a baseline for evaluating how well different aggregation methods recover the ground truth. In \Cref{sec:benchmark}, where we apply both traditional voting rules and SP-Voting to aggregate responses, this baseline enables us to quantify the impact of incorporating meta-information and to measure the improvement in performance achieved by the SP-Voting approach.

\section{\texttt{SP-Rank} Benchmark}
\label{sec:benchmark}
Benchmarking is critical for evaluating the reliability of rank aggregation methods in crowdsourced settings, where individual preferences are noisy and voter quality varies. Traditional voting rules operate purely on first-order information—participants’ votes—and typically assume uniform reliability across individuals. \texttt{SP-Rank} supports these classical aggregation methods while also enabling the application of second-order approaches like SP-Voting, which incorporate each voter’s prediction of others’ preferences to better infer the ground-truth answer. This dual structure allows researchers to compare and evaluate a broad spectrum of aggregation strategies under a common framework.

By benchmarking SP-Voting against classical aggregation rules—Borda \cite{borda1781m} (which assigns points based on ranking positions), Copeland \cite{copeland1951reasonable} (which scores candidates by head-to-head victories), and Maximin \cite{young1977extending} (which considers the strongest worst-case pairwise performance)—we quantify the added value of incorporating second-order information. This evaluation also establishes a reference point for future methods that leverage meta-predictions, enabling comparison not only with traditional baselines but also with a principled second-order approach.

We evaluate \texttt{SP-Rank} across the following key tasks - 1) Full ground-truth rank recovery, 2) Subset-level ground-truth rank recovery 3) Preference model accuracy (Learning to Rank)

\begin{algorithm}[!h]
\caption{\textbf{Partial-SP for Full Ground-Truth Rank Recovery}}
\label{alg:partial-sp-full}
\DontPrintSemicolon
\SetKwInput{Input}{Inputs}
\SetKwInput{Output}{Output}
\Input{
Dataset $\mathcal{D}$ of crowdsourced responses partitioned by domain $d\in\{\text{Geo},\text{Movies},\text{Paintings}\}$; \\
for each problem $p$ and subset $S\subseteq A_d$ with $|S|=k\in\{4,5\}$: voter \emph{votes} and \emph{predictions} collected under an elicitation format $t$;
ground-truth ranking $\sigma_d^\star$ over $M$ items per domain $M\in\{50,36\}$, and a choice of final aggregation rule $R\in\{\text{Borda},\text{Copeland},\text{Maximin}\}$.
}
\Output{Estimated full ranking $\hat{\sigma}_d$ over $A_d$ and Kendall-Tau distance $\tau_K(\hat{\sigma}_d,\sigma_d^\star)$.}

\BlankLine
\textbf{(A) Local SP estimation on each subset}\;
\ForEach{domain $d$}{
  Initialize empty multiset $\mathcal{P}_d$ of partial orders.\;
  \ForEach{problem $p$ in domain $d$}{
    \ForEach{subset $S\subseteq A_d$ shown in $p$}{
      \tcp{Compute SP over \emph{partial} rankings on $S$ from votes + predictions}
      Estimate empirical vote frequencies $f_S(\sigma)$ over partial orders $\sigma$ on $S$ (per elicitation $t$).\;
      From predictions, estimate cross-probabilities $g_S(\sigma'\mid \sigma)$ between partial orders on $S$.\;
      Define Prediction-normalized score $V_S(\sigma) \leftarrow f_S(\sigma)\cdot \sum_{\sigma'}\frac{g_S(\sigma'\mid \sigma)}{g_S(\sigma\mid \sigma')}$.\;
      Select $\hat{\sigma}_S \leftarrow \arg\max_{\sigma} V_S(\sigma)$ (break ties uniformly at random).\;
      Insert $\hat{\sigma}_S$ into $\mathcal{P}_d$.\;
    }
  }
}

\BlankLine
\textbf{(B) Lift partial orders to pairwise tallies}\;
\ForEach{domain $d$}{
  Initialize pairwise wins $W_d(a,b)\leftarrow 0$ and comparisons $C_d(a,b)\leftarrow 0$ for all distinct $a,b\in A_d$.\;
  \ForEach{$\hat{\sigma}_S\in\mathcal{P}_d$}{
    \ForEach{ordered pair $(a,b)\in S\times S,\, a\neq b$}{
      \If{$a \succ_{\hat{\sigma}_S} b$}{ $W_d(a,b)\leftarrow W_d(a,b)+1$; $C_d(a,b)\leftarrow C_d(a,b)+1$; $C_d(b,a)\leftarrow C_d(b,a)+1$ }
    }
  }
  Define support $P_d(a,b)\leftarrow \frac{W_d(a,b)}{\max\{1,\,C_d(a,b)\}}$ for all $a\neq b$.\;
}

\BlankLine
\textbf{(C) Aggregate to a full ranking with rule $R$}\;
\ForEach{domain $d$}{
  \eIf{$R=\text{Copeland}$}{
    \ForEach{$a\in A_d$}{ $\text{score}(a)\leftarrow \sum_{b\neq a}\mathbf{1}\{P_d(a,b)>P_d(b,a)\}-\mathbf{1}\{P_d(a,b)<P_d(b,a)\}$ }
    $\hat{\sigma}_d \leftarrow$ items sorted by $\text{score}$ (ties $\to$ random).\;
  }{
    \eIf{$R=\text{Maximin}$}{
      \ForEach{$a\in A_d$}{ $\text{score}(a)\leftarrow \min_{b\neq a} P_d(a,b)$ }
      $\hat{\sigma}_d \leftarrow$ items sorted by $\text{score}$ (ties $\to$ random).\;
    }{
      \tcp{$R=\text{Borda}$ via pairwise-approx. positional scoring}
      \ForEach{$a\in A_d$}{ $\text{score}(a)\leftarrow \sum_{b\neq a}\big(P_d(a,b)-P_d(b,a)\big)$ }
      $\hat{\sigma}_d \leftarrow$ items sorted by $\text{score}$ (ties $\to$ random).\;
    }
  }
}

\BlankLine
\textbf{(D) Evaluation}\;
\ForEach{domain $d$}{
  Compute Kendall’s $\tau_K(\hat{\sigma}_d,\sigma_d^\star)$ over all $M$ items.\;
}
\Return{$\{\hat{\sigma}_d,\tau_K(\hat{\sigma}_d,\sigma_d^\star)\}_{d}$}\;

\BlankLine
\end{algorithm}
\subsection{Full Ground-Truth Rank Recovery}

%\HH{We should also briefly define the voting rules (Borda, Copeland, etc.) and add their detailed and formal definitions in the appendix (could be just bringing these from our other paper with some paraphrasing).}

This task evaluates how accurately the full ground-truth ranking can be reconstructed by aggregating voter preferences (which are provided at a subset level). We analyze both the 4-alternative and 5-alternative conditions, which correspond to underlying ground-truth rankings over 50 and 36 total alternatives, respectively. Notably, the first 720 respondents were assigned a ground-truth ranking with 50 alternatives, while the remaining participants received a ground-truth ranking with 36 alternatives. As a result, for full ground-truth recovery tasks, we treat these as two distinct datasets. The Partial-SP  variant of SP-Voting (as shown in \Cref{alg:partial-sp-full}) proposed by \citet{hosseini2025surprisingly} is benchmarked against Borda, Copeland, and Maximin in \Cref{tab:full_groundtruth_kendall_tau} and \Cref{tab:domain_tau_full_recovery_combined}.

\begin{table*}[ht]
\centering
\begin{tabular}{lcccccc}
\toprule
\textbf{Elicitation Format} & \multicolumn{2}{c}{Borda} & \multicolumn{2}{c}{Copeland} & \multicolumn{2}{c}{Maximin} \\
\cmidrule(lr){2-3} \cmidrule(lr){4-5} \cmidrule(lr){6-7}
\textbf{Subset Size = 4 (50 alternatives)} & Vote & SP & Vote & SP & Vote & SP \\
\midrule
Rank-Rank & 0.08 & \textbf{0.40} & 0.13 & \textbf{0.54} & 0.40 & \textbf{0.62} \\
Rank-App1 & 0.08 & \textbf{0.27} & 0.15 & \textbf{0.34} & 0.39 & \textbf{0.45} \\
Top-Rank & 0.11 & \textbf{0.27} & 0.11 & \textbf{0.33} & 0.42 & \textbf{0.48} \\
Top-Top  & 0.11 & \textbf{0.21} & 0.11 & \textbf{0.26} & 0.41 & \textbf{0.38} \\
\midrule
\textbf{Subset Size = 5 (36 alternatives)} & \multicolumn{2}{c}{Borda} & \multicolumn{2}{c}{Copeland} & \multicolumn{2}{c}{Maximin} \\
\cmidrule(lr){2-3} \cmidrule(lr){4-5} \cmidrule(lr){6-7}
App2-App2     & 0.08 & \textbf{0.62} & 0.09 & \textbf{0.72} & 0.40 & \textbf{0.81} \\
App3-Rank     & 0.02 & \textbf{0.70} & 0.08 & \textbf{0.85} & 0.41 & \textbf{0.92} \\
Rank-Rank     & 0.03 & \textbf{0.67} & 0.09 & \textbf{0.81} & 0.40 & \textbf{0.86} \\
Rank-App1     & 0.05 & \textbf{0.41} & 0.11 & \textbf{0.44} & 0.41 & \textbf{0.53} \\
App1-App3     & 0.04 & \textbf{0.28} & 0.05 & \textbf{0.28} & \textbf{0.40} & \textbf{0.40} \\
App1-Rank      & 0.06 & \textbf{0.46} & 0.05 & \textbf{0.52} & 0.36 & \textbf{0.61} \\
App1-App1       & 0.11 & \textbf{0.27} & 0.11 & \textbf{0.25} & \textbf{0.40} & 0.38 \\
\bottomrule
\end{tabular}
\caption{Average Kendall Tau correlations (Vote vs SP) for full ground-truth rank recovery across domains. Results are grouped by subset size that voters voted on and ground truth ranking size.}
\label{tab:full_groundtruth_kendall_tau}
\end{table*}

As shown in \Cref{tab:full_groundtruth_kendall_tau}, SP-Voting consistently outperforms traditional vote-based aggregation across all elicitation formats and voting rules for both the 4-alternative and 5-alternative conditions. In the 4-alternative setting with 50 ground-truth items, SP-Voting yields substantial gains in Kendall Tau correlation, particularly under Copeland and Maximin (e.g., \texttt{Rank-Rank}: 0.54 vs.\ 0.13 and 0.62 vs.\ 0.40, respectively). Even in limited-information formats such as \texttt{App1-App1}, SP-Voting provides meaningful improvements. The advantage becomes even more pronounced in the 5-alternative setting with 36 ground-truth items. Here, SP-Voting achieves large absolute improvements across nearly all format pairs, with the \texttt{App3-Rank} condition improving from 0.08 to 0.85 under Copeland and from 0.41 to 0.92 under Maximin. These results also suggest that eliciting preferences over larger subsets brings us closer to the ground truth, as voters provide more informative signals when reasoning over a broader set of alternatives.Notably however, SP-Voting also performs well even when the elicited information is minimal (for example, in \texttt{App2-App2}), indicating its robustness in leveraging second-order beliefs. All reported results are averaged across three domains—geography, movies, and paintings.

\begin{table*}[ht]
\centering
\begin{tabular}{lcccccc}
\toprule
\textbf{Elicitation Format} & \multicolumn{2}{c}{Geography} & \multicolumn{2}{c}{Movies} & \multicolumn{2}{c}{Paintings} \\
\cmidrule(lr){2-3} \cmidrule(lr){4-5} \cmidrule(lr){6-7}
\textbf{Subset Size = 4 (50 alternatives)} & Vote & SP & Vote & SP & Vote & SP \\
\midrule
Rank-Rank     & 0.26 & \textbf{0.55} & 0.20 & \textbf{0.55} & 0.14 & \textbf{0.46} \\
Rank-App1     & 0.29 & \textbf{0.43} & 0.20 & \textbf{0.37} & 0.12 & \textbf{0.27} \\
App1-Rank      & 0.34 & \textbf{0.38} & 0.17 & \textbf{0.33} & 0.14 & \textbf{0.37} \\
App1-App1       & 0.31 & \textbf{0.34} & 0.20 & \textbf{0.28} & 0.14 & \textbf{0.23} \\
\midrule
\textbf{Subset Size = 5 (36 alternatives)} & \multicolumn{6}{c}{} \\
\cmidrule(lr){1-7}
App2-App2     & 0.28 & \textbf{0.75} & 0.15 & \textbf{0.72} & 0.13 & \textbf{0.68} \\
App3-Rank     & 0.20 & \textbf{0.82} & 0.17 & \textbf{0.81} & 0.14 & \textbf{0.84} \\
Rank-Rank     & 0.25 & \textbf{0.77} & 0.07 & \textbf{0.74} & 0.20 & \textbf{0.84} \\
Rank-App1     & 0.24 & \textbf{0.56} & 0.17 & \textbf{0.42} & 0.16 & \textbf{0.39} \\
App1-App3     & 0.30 & \textbf{0.40} & 0.13 & \textbf{0.28} & 0.06 & \textbf{0.28} \\
App1-Rank      & 0.15 & \textbf{0.55} & 0.14 & \textbf{0.49} & 0.18 & \textbf{0.56} \\
App1-App1       & 0.26 & \textbf{0.36} & 0.16 & \textbf{0.29} & 0.19 & \textbf{0.24} \\
\bottomrule
\end{tabular}
\caption{Average Domain-wise Kendall Tau correlations (Vote vs SP) for complete ground-truth rank recovery. Results are grouped by subset size that voters voted on and ground truth ranking size.}
\label{tab:domain_tau_full_recovery_combined}
\end{table*}

As shown in \Cref{tab:domain_tau_full_recovery_combined}, SP-Voting consistently improves full ground-truth rank recovery across all domains—geography, movies, and paintings—for both subset sizes. In the 4-alternative setting with 50 total items, SP-Voting yields moderate but consistent gains in Kendall Tau correlations across all elicitation format pairs. For example, in the \texttt{Rank-Rank} condition, SP-Voting improves from 0.14 to 0.46 in paintings, indicating its ability to enhance signal even in low-agreement domains. In the 5-alternative setting with 36 items, improvements are more substantial and robust. Notably, SP-Voting achieves gains of over 0.5 points in some cases, such as in \texttt{App3-Rank} (e.g., 0.20 to 0.82 in geography and 0.14 to 0.84 in paintings), and consistently performs well even in elicitation formats with sparse or noisy vote signals (e.g., \texttt{App2-App2}, \texttt{App1-Rank}). These results demonstrate that the benefit of incorporating second-order information generalizes across domains with varying difficulty and voter agreement.

\subsection{Subset Level Ground-Truth Rank Recovery}

 Here, we assess how well the correct ranking is recovered within each individual subset of alternatives shown to participants. Each unique subset is evaluated independently, allowing a fine-grained comparison of Partial-SP and traditional methods at a local level.

\begin{table*}[h]
\centering

\begin{tabular}{lcccccc}
\toprule
\textbf{Elicitation Format} & \multicolumn{2}{c}{Borda} & \multicolumn{2}{c}{Copeland} & \multicolumn{2}{c}{Maximin} \\
\cmidrule(lr){2-3} \cmidrule(lr){4-5} \cmidrule(lr){6-7}
\textbf{Subset Size = 4 (Subset Level Ground Truth)} & Vote & SP & Vote & SP & Vote & SP \\
\midrule
Rank-Rank     & 0.14 & \textbf{0.65} & 0.15 & \textbf{0.66} & 0.11 & \textbf{0.65} \\
Rank-App1     & 0.14 & \textbf{0.31} & 0.23 & \textbf{0.39} & 0.10 & \textbf{0.31} \\
App1-Rank      & 0.20 & \textbf{0.61} & 0.20 & \textbf{0.61} & 0.13 & \textbf{0.61} \\
App1-App1       & 0.17 & \textbf{0.32} & 0.17 & \textbf{0.32} & 0.11 & \textbf{0.32} \\
\midrule
\textbf{Subset Size = 5 (Subset Level Ground Truth)} & \multicolumn{6}{c}{} \\
\cmidrule(lr){1-7}
App2-App2     & 0.09 & \textbf{0.72} & 0.08 & \textbf{0.75} & 0.05 & \textbf{0.72} \\
App3-Rank     & 0.02 & \textbf{0.86} & 0.04 & \textbf{0.88} & 0.03 & \textbf{0.86} \\
Rank-Rank     & 0.07 & \textbf{0.83} & 0.03 & \textbf{0.85} & 0.04 & \textbf{0.83} \\
Rank-App1     & 0.10 & \textbf{0.40} & 0.09 & \textbf{0.40} & 0.07 & \textbf{0.40} \\
App1-App3     & 0.05 & \textbf{0.30} & 0.05 & \textbf{0.30} & 0.02 & \textbf{0.30} \\
App1-Rank      & 0.05 & \textbf{0.57} & 0.05 & \textbf{0.57} & 0.03 & \textbf{0.57} \\
App1-App1       & 0.10 & \textbf{0.24} & 0.10 & \textbf{0.24} & 0.06 & \textbf{0.24} \\
\bottomrule
\end{tabular}
\caption{Average Subset-level Kendall Tau correlations (Vote vs SP) for ground-truth rank recovery \textit{within subsets} across domains. Results are grouped by subset size that voters voted on.}
\label{tab:subset_tau_partial_recovery_combined}
\end{table*}

As shown in \Cref{tab:subset_tau_partial_recovery_combined}, SP-Voting substantially outperforms traditional vote-based aggregation in recovering the correct ranking within each subset of alternatives. In the 4-alternative setting, SP-Voting shows higher Kendall Tau correlations across all elicitation formats and voting rules, with particularly strong gains observed in \texttt{Rank-Rank} and \texttt{App1-Rank} (e.g., Maximin: 0.11 to 0.65 and 0.13 to 0.61, respectively). Even in lower-information formats such as \texttt{App1-App1}, SP-Voting offers consistent improvements over vote-only baselines. The 5-alternative setting shows even greater performance gains. For example, in the \texttt{App3-Rank} condition, Kendall Tau increases from 0.02 to 0.86 under Borda, and from 0.03 to 0.86 under Maximin—highlighting the effectiveness of SP-Voting even when individual votes are weakly informative. Across all subset sizes and elicitation formats, SP-Voting demonstrates robust improvements, emphasizing the strength of second-order signals in enabling accurate local rank recovery.

\begin{table*}[h]
\centering
\begin{tabular}{lcccccc}
\toprule
\textbf{Elicitation Format} & \multicolumn{2}{c}{Geography} & \multicolumn{2}{c}{Movies} & \multicolumn{2}{c}{Paintings} \\
\cmidrule(lr){2-3} \cmidrule(lr){4-5} \cmidrule(lr){6-7}
\textbf{Subset Size = 4 (Subset Level Ground Truth)} & Vote & SP & Vote & SP & Vote & SP \\
\midrule
Rank-Rank       & 0.35 & \textbf{0.72} & 0.15 & \textbf{0.69} & -0.10 & \textbf{0.56} \\
Rank-App1       & 0.40 & \textbf{0.46} & 0.14 & \textbf{0.37} & -0.07 & \textbf{0.18} \\
App1-Rank        & 0.32 & \textbf{0.64} & 0.15 & \textbf{0.60} &  0.05 & \textbf{0.59} \\
App1-App1         & 0.33 & \textbf{0.47} & 0.14 & \textbf{0.36} & -0.03 & \textbf{0.13} \\
\midrule
\textbf{Subset Size = 5 (Subset Level Ground Truth)} & \multicolumn{2}{c}{Geography} & \multicolumn{2}{c}{Movies} & \multicolumn{2}{c}{Paintings} \\
\cmidrule(lr){2-3} \cmidrule(lr){4-5} \cmidrule(lr){6-7}
\textbf{Elicitation Format} & Vote & SP & Vote & SP & Vote & SP \\
\midrule
App2-App2       & 0.21 & \textbf{0.76} & 0.03 & \textbf{0.74} & -0.02 & \textbf{0.69} \\
App3-Rank       & 0.07 & \textbf{0.86} & 0.03 & \textbf{0.86} & -0.01 & \textbf{0.89} \\
Rank-Rank       & 0.14 & \textbf{0.84} & -0.08 & \textbf{0.77} & 0.09 & \textbf{0.91} \\
Rank-App1       & 0.20 & \textbf{0.52} & 0.05 & \textbf{0.38} & 0.03 & \textbf{0.31} \\
App1-App3        & 0.22 & \textbf{0.46} & 0.01 & \textbf{0.22} & -0.11 & \textbf{0.22} \\
App1-Rank        & 0.11 & \textbf{0.61} & -0.02 & \textbf{0.51} & 0.05 & \textbf{0.60} \\
App1-App1         & 0.16 & \textbf{0.27} & 0.03 & \textbf{0.24} & 0.06 & \textbf{0.22} \\
\bottomrule
\end{tabular}
\caption{Average Domain-wise Kendall Tau correlations (Vote vs SP) for subset-level ground-truth recovery \textit{within
subsets}. Results are grouped by subset size that voters voted on.}
\label{tab:domain-kendall-tau-subset}
\end{table*}

As shown in \Cref{tab:domain-kendall-tau-subset}, SP-Voting consistently improves subset-level ground-truth rank recovery across all domains and elicitation formats for both the 4-alternative and 5-alternative settings. In the 4-alternative condition, SP-Voting yields substantial gains, particularly in \texttt{Rank-Rank}, where Kendall Tau increases from 0.35 to 0.72 in geography and from -0.10 to 0.56 in paintings—demonstrating strong recovery even when vote-only baselines are weakly or negatively aligned with ground truth. The 5-alternative condition exhibits even larger improvements, with SP-Voting reaching Tau scores above 0.85 in \texttt{App3-Rank} across all domains and surpassing 0.90 in \texttt{Rank-Rank} for paintings. Interestingly, the paintings domain—being the most subjective of the three—shows some of the weakest baseline correlations, yet the most dramatic gains from SP-Voting (e.g., \texttt{App2-App2}: -0.02 to 0.69, \texttt{App1-App3}: -0.11 to 0.22). This highlights SP-Voting’s ability to leverage second-order information to recover ground-truth, even in domains with high variability in individual preferences.

\subsection{Preference Model Accuracy (Learning to Rank)}

In this task, we evaluate probabilistic models based on their ability to replicate real-world voter behavior. We fit models to elicited preferences in real data, simulate synthetic votes using the learned parameters, and then refit the model to the synthetic data. Predictive accuracy is assessed by computing the relative error between parameters inferred from real and synthetic data, where relative error is defined as the absolute difference between parameter estimates divided by the original estimate. The modeling approach uses the framework introduced in \citet{hosseini2025surprisingly}.

\begin{table*}[h!]
\centering
\begin{tabular}{llll}
\toprule
\textbf{Model} & \textbf{Setting} & \textbf{Parameter} & \textbf{Relative Error} \\
\midrule
\multirow{5}{*}{Mallows} 
  & 2-group & Expert Dispersion (vote) $\phi_{ev}$ & 0.03 \\
  &         & Expert Dispersion (prediction) $\phi_{ep}$ & 0.01 \\
  &         & Non-expert Dispersion (vote) $\phi_{nev}$ & 0.37 \\
  &         & Non-expert Dispersion (prediction) $\phi_{nep}$ & 0.49 \\
  &         & Proportion of Experts $\pi_e$ & 0.02 \\
\midrule
\multirow{5}{*}{Mallows} 
  & 3-group & Expert Dispersion (vote) $\phi_{ev}$ & 0.80 \\
  &         & Expert Dispersion (prediction) $\phi_{ep}$ & 0.94 \\
  &         & Non-expert Dispersion (vote) $\phi_{nev}$ & 0.36 \\
  &         & Non-expert Dispersion (prediction) $\phi_{nep}$ & 0.51 \\
  &         & Proportion of Experts $\pi_e$ & 0.37 \\
\midrule
\multirow{5}{*}{Plackett-Luce} 
  & 2-group & Expert Strength (vote) $s_{ev}$ & 0.18 \\
  &         & Expert Strength (prediction) $s_{ep}$ & 0.94 \\
  &         & Non-expert Strength (vote) $s_{nev}$ & 0.31 \\
  &         & Non-expert Strength (prediction) $s_{nep}$ & 1.54 \\
  &         & Proportion of Experts $\pi_e$ & 0.69 \\
\midrule
\multirow{5}{*}{Plackett-Luce} 
  & 3-group & Expert Strength (vote) $s_{ev}$ & 0.17 \\
  &         & Expert Strength (prediction) $s_{ep}$ & 0.78 \\
  &         & Non-expert Strength (vote) $s_{nev}$ & 0.93 \\
  &         & Non-expert Strength (prediction) $s_{nep}$ & 1.98 \\
  &         & Proportion of Experts $\pi_e$ & 0.41 \\
\bottomrule
\end{tabular}
\caption{Relative Error Comparison of Inferred Parameters from Votes and Predictions for Mallows and Plackett-Luce Models}
\label{tab:relative_error_detailed}
\end{table*}

As shown in \Cref{tab:relative_error_detailed}, probabilistic models demonstrate varying degrees of success in recovering parameters from elicited preferences. In the 2-group Mallows model, expert dispersion parameters are recovered with high accuracy from both votes and predictions ($\phi_{ev} = 0.03$, $\phi_{ep} = 0.01$), while errors are notably higher for non-expert parameters, especially predictions ($\phi_{nep} = 0.49$). The 3-group Mallows setting yields a similar pattern but with degraded expert recovery ($\phi_{ev} = 0.80$, $\phi_{ep} = 0.94$), suggesting increased model complexity reduces reliability when inferring fine-grained group structures. The Plackett-Luce model shows larger relative errors overall, particularly for prediction-based parameters in the non-expert group ($s_{nep} = 1.54$ in 2-group, $1.98$ in 3-group). Across both models, the proportion of experts $\pi_e$ is recovered more reliably in simpler settings, with errors increasing in 3-group configurations. These results indicate that while preference models can effectively capture behavior of different sub-groups in the population.

\section{Discussion}

This work introduces \texttt{SP-Rank}, the first large-scale dataset to support the evaluation of both traditional and second-order aggregation methods in realistic crowdsourced settings. With over 12,000 human-generated datapoints spanning diverse domains and elicitation formats—and including both individual votes and meta-predictions—\texttt{SP-Rank} enables rigorous benchmarking under noisy, heterogeneous conditions. Our results show that SP-Voting consistently outperforms classical vote-only methods such as Borda, Copeland, and Maximin, including in low-information formats and subjective domains, underscoring the value of second-order information. \texttt{SP-Rank} also enables the training and evaluation of probabilistic models that capture structured group behavior and latent expertise. However, the dataset has certain limitations. It covers only three domains—geography, movies, and paintings—which may not represent the full spectrum of real-world ranking tasks. Additionally, participants were drawn exclusively from Amazon Mechanical Turk, with regional restrictions that may limit the generalizability of findings across cultural or demographic contexts. Finally, we assume the existence of a single objective or subjective ground truth for each question, abstracting away from real-world cases where consensus may be ambiguous or contested. These design choices reflect the absence of existing real-world datasets with second-order signals. We view this release as a foundational step and invite the community to extend SP-Rank into new domains and cultural contexts.

\texttt{SP-Rank} opens up several promising directions for future research. Its dual-layer structure—combining first- and second-order information—provides a foundation for studying belief formation, expertise inference, and aggregation under uncertainty. The dataset can be used to evaluate not only classical voting rules and SP-based methods but also emerging neural or LLM-based aggregation techniques, facilitating cross-method comparisons. It is also well-suited for training reward models in preference-based learning pipelines such as Reinforcement Learning from Human Feedback (RLHF) and Direct Preference Optimization (DPO). Future expansions of \texttt{SP-Rank} could incorporate additional domains, multilingual or cross-cultural populations, and dynamic or time-sensitive preferences, further broadening its utility in understanding human judgments and improving AI alignment. By introducing SP-Rank, we aim not only to benchmark current aggregation methods but also to catalyze broader data collection efforts. Expanding the availability of datasets with joint vote–prediction data will be essential for advancing preference aggregation, human–AI alignment, and downstream applications.

\newpage

%\subsubsection*{Acknowledgments}

\newpage
\printbibliography

% \begin{acks}
% \end{acks}

\newpage

\appendix
\clearpage
\onecolumn
    \end{document}